\documentclass[apj]{emulateapj}
\usepackage{apjfonts,mathptmx}

\def\cxo {\emph{Chandra}}

\begin{document}
\journalinfo{The Astrophysical Journal, in press}
\submitted{The Astrophysical Journal, in press}

\title{\emph{Chandra} astrometry sets a tight upper limit to the\\
proper motion of SGR\,1900+14}

\author{A.~De~Luca,\altaffilmark{1,2,4} P.~A.~Caraveo,\altaffilmark{1} P.~Esposito,\altaffilmark{1,3,4} and K.~Hurley\altaffilmark{5}}

\altaffiltext{1}{INAF - Istituto di Astrofisica Spaziale e Fisica Cosmica Milano, via E.~Bassini 15, 20133 Milano, Italy}
\altaffiltext{2}{IUSS - Istituto Universitario di Studi Superiori, viale Lungo Ticino Sforza, 56, 27100 Pavia, Italy}
\altaffiltext{3}{Universit\`a degli Studi di Pavia, Dipartimento di Fisica Nucleare e Teorica, via A.~Bassi 6, 27100 Pavia, Italy}
\altaffiltext{4}{INFN - Istituto Nazionale di Fisica Nucleare, sezione di Pavia, via A.~Bassi 6, 27100 Pavia, Italy}
\altaffiltext{5}{Space Sciences Laboratory, University of California, 7 Gauss Way Berkeley, CA 94720-7450 USA}
\email{deluca@iasf-milano.inaf.it}

\shortauthors{A.~De~Luca et al.}
\shorttitle{\emph{Chandra} proper motion of SGR\,1900+14}


\begin{abstract}
The soft gamma-ray repeater (SGR) SGR\,1900+14 lies a few arcminutes
outside the edge of the shell supernova remnant (SNR) G42.8+0.6. A 
physical association between the two systems has been proposed - for 
this and other SGR-SNR pairs - based on the expectation of high space 
velocities for SGRs in the framework of the magnetar model. The large 
angular separation between the SGR and the SNR center, coupled with the
young age of the system, suggest a test of the association with a proper
motion measurement.
We used a set of three \cxo/ACIS 
observations of the field spanning $\sim5$ years to perform accurate 
relative astrometry in order to measure the possible angular displacement 
of the SGR as a function of time. Our investigation sets a $3\sigma$ upper 
limit of 70 mas yr$^{-1}$ to the overall proper motion of the SGR. Such a 
value argues against an association of SGR\,1900+14 with G42.8+0.6 and adds 
further support to the mounting evidence for an origin of the SGR within a 
nearby, compact cluster of massive stars.
\end{abstract}

\keywords{X-rays: individual (SGR\,1900+14) -- ISM: individual (G42.8+0.6) -- X-rays: stars -- stars: neutron -- supernova remnants.}

\section{Introduction}

Soft gamma-ray repeaters 
\citep[SGRs - see e.g. ][for reviews]{hurley00sgr,woods06,mereghetti08}
are a handful of four (plus a few candidates) sources of short 
bursts of soft gamma rays. SGRs were originally thought to be a peculiar 
subclass of Gamma-Ray Bursts (GRBs). 
Contrary to the behavior of  classical GRBs, SGRs produce series of bursts 
over various time scales and the events are characterized by a soft spectral 
shape. SGRs have a very rich high energy phenomenology. Apart from the flaring 
activity, characterized by the emission of multiple, short ($\sim$0.1 s) 
bursts with peak luminosities of $\sim$$10^{41}$ erg~s$^{-1}$, sometimes 
culminating in dramatic, very energetic \emph{giant flares} 
with peak luminosity exceeding $10^{47}$ erg~s$^{-1}$ 
\citep[as was the case of the 2004 December 27 event from SGR\,1806--20 -][]{hurley05short}, 
SGRs display persistent emission from 0.1 to hundreds of keV, with pulsations 
in the 5--8 s range, and with significant variability in flux, spectral shape, 
pulse shape and pulsed fraction; they also exhibit glitches and an 
irregular period derivative.

This phenomenology has been  interpreted within the framework of the
\emph{magnetar} model. SGRs are believed to be young, isolated neutron stars
(INSs) endowed with an ultra-high magnetic field (\mbox{$B\sim10^{15}$ G}), 
which is thought to be the energy reservoir for all high energy emissions 
\citep{duncan92,thompson95,thompson96}. It is now commonly accepted that 
Anomalous X-ray Pulsars (AXPs), a peculiar class of X-ray pulsars 
\citep[see ][for reviews]{woods06,mereghetti08}, could also be magnetars, 
in view of a close similarity of several aspects of the high energy 
phenomenology.\\
\indent The identification of SGRs with young INSs is on a rather firm basis. 
However, in the past, an important supporting argument has been their possible 
association with supernova remnants (SNRs). This provided a link between SGR' 
formation and supernova explosions, and also set an upper limit to their age, 
in view of the short lives of SNRs. Indeed, a possible SNR association had been
claimed for all SGRs \citep{hurley00sgr}. The validity of this association was
later reconsidered as new multiwavelength observations became available 
\citep[see ][and references therein]{gaensler01}.
In one case (SGR\,1806--20) the mere existence of a nearby SNR was ruled out.
For the remaining cases, the SGR positions, close to the edges or even outside
their SNRs, cast doubt on the associations, based on two main considerations: 
(i) the significant chance alignement probability; (ii) the need for large 
spatial velocities (well in excess of 1000 km s$^{-1}$). 
Although such large velocities are not unheard of in the neutron star family 
\citep{hobbs05}, AXPs do not show any evidence for them. This would suggest 
that AXPs and SGRs are different classes of INSs, originating in intrinsically 
different supernova explosion processes, in contrast to the robust evidence 
for a very close relationship between SGR and AXP families. In view of this, 
all SGR-SNR associations have been reconsidered, and a possible association of 
two SGRs with nearby, massive star clusters has been proposed (SGR\,1806--20 
by \citealt{fuchs99}; SGR\,1900+14 by \citealt{vrba00}).



\section{SGR\,1900+14 and SNR G42.8+0.6}
SGR\,1900+14 lies outside the rim of G42.8+0.6, a shell-type, $\sim$$10^4$ 
year old radio SNR located at a distance of $3\div9$ kpc \citep{mlrh01}. An 
association between the two systems was originally proposed by 
\citet{vasisht94} and \citet{hurley96}.
SGR\,1900+14 lies towards a rather complicated region of the Galaxy, and 
\citet{gaensler01} estimated a chance probability for the alignment of a 
SNR to be as high as 4\%. \citet{kaplan02sgr} even increased this estimate 
to $\sim$23\%. \citet{lorimer00} detected a young ($\tau_c \sim 38$ kyr) 
radio pulsar about 2 arcmin away from the SGR. Such a pulsar could also be 
plausibly associated with G42.8+0.6. Moreover, \citet{vrba00} discovered
a massive star cluster very close ($\sim$12 arcsec) to the position of the 
SGR. Such observations weakened the case for an association between SGR\,1900+14 
and G42.8+0.6. Very recently, \citet{wachter08} discovered an infrared ring 
surrounding SGR\,1900+14 with \emph{Spitzer}. This structure was interpreted 
as the rim of a dust cavity produced by past giant flares from the SGR, 
heated by a nearby star cluster. This would point to a possible association 
of SGR\,1900+14 with the star cluster, although a large difference in 
reddening towards the SGR and the cluster remains to be explained

In this work, we directly probe the association of SGR\,1900+14 and the SNR 
G42.8+0.6 through a proper motion measurement. Such a possibility had been 
suggested by \citet{hurley02}.
The angular separation between the SGR and the center of the SNR is $\sim$18 
arcmin \citep{hurley02}. 
Adopting a value of \mbox{$10^4$ yr} 
as a conservative estimate for the age of the system
\citep{thompson00}, the SGR-SNR association would 
require  a proper motion of at least 0.11 arcsec yr$^{-1}$.
Such a proper motion can be measured using multi-epoch observations with the 
\cxo\ X-ray Observatory. The superb angular resolution of its optics and 
the stability of aspect reconstruction with its imaging detectors allow 
measurements of tiny angular displacements through accurate relative 
astrometry \citep[see e.g.][]{motch07,motch08}.

\section{\emph{Chandra} observations and data analysis}
\label{data}

SGR\,1900+14 has been observed three times by \cxo\ between 2001 and 2006
using the Advanced CCD Imaging Spectrometer (ACIS). The first observation 
was carried out in response to an AO-2 proposal whose goal was to obtain a 
baseline measurement for proper motion studies. A log of the available data 
is given in Table~\ref{cxodata}.
\begin{deluxetable}{lcccc}
  \tablecolumns{1}
\tablewidth{0pt}
\tablecaption{\emph{Chandra} observations of the field of SGR\,1900+14.\label{cxodata}}
\tablehead{
\colhead{Date} & \colhead{MJD} & \colhead{Obs. ID} & \colhead{Instrument} & \colhead{Exposure time}
}
\startdata
2001 Jun 17 & 52,077 & 1954 & ACIS/I  & 30.1 ks\\
2002 Mar 11 & 52,344 & 3449 & ACIS/S & 2.7 ks \\
2006 Jun 4 & 53,890 &  6731 & ACIS/I  & 25.0 ks\enddata
\tablecomments{All observations were performed using the Timed Exposure mode and the Faint event telemetry format.}
\end{deluxetable}
We retrieved event files from the \cxo\ X-ray Center Data Archive. All the 
datasets have gone through 
`reprocessing III'\footnote{See http://cxc.harvard.edu/ciao/repro\_iii.html.}
with updated software and calibration. According to the \cxo\ X-ray Center 
guidelines,\footnote{See http://cxc.harvard.edu/ciao/threads/createL2/.}
no further reprocessing is required and archival `level 2' data were adopted 
as a starting point. We used the \cxo\ Interactive Analysis of Observation 
software (CIAO version 3.3) for our analysis.

No significant background flares affected the observations. We removed pixel 
randomization (we checked \emph{a posteriori} that fully consistent results 
are obtained by using standard pixel position randomization) and we
generated images of the field by selecting photons in the 0.3--8 keV energy 
range, binning the CCD pixel size by a factor of 2. The target was imaged 
close to the aimpoint on the ACIS/I3 detector for Observations 1 and 3 and 
on the ACIS/S3 detector for Observation 2.

Source detection was done using the WAVDETECT task, with wavelet 
scales ranging from 1 to 16 pixels, spaced by a factor $\sqrt{2}$. A 
detection threshold of $10^{-5}$ was selected in order to avoid missing 
faint sources. Cross-correlation of the source lists produced for each 
observation (adopting a maximum source distance of 3 arcsec) allowed us to 
reject spurious detections (about $\sim$70 per field in Observation 1 and 
Observation 3). We identified a set of 31 common sources in Observation 1 
and Observation 3 (excluding the target), while only six such sources 
were found in Observation 2, which was significantly shorter and had a 
smaller field of view.

The probability of a chance alignment of two false detections is estimated 
to be of order $\leq$0.1\%. Thus, we may safely assume that all of the 
selected common sources are real.

The uncertainty affecting the source's positions in the reference frame of 
each image depends on the source signal-to-noise, as well as on the distance
from the aimpoint (because of point-spread function [PSF] degradation as a 
function of off-axis angle). The position of the target, the brightest source
in the field, located close to the aimpoint, was determined with a $1\sigma$ 
error of 0.02 pixels per coordinate, while the position of a typical faint 
source located several arcmin off-axis is affected by an uncertainty of order 
1 pixel per coordinate.

\section{Relative Astrometry}

Relative astrometry relies on accurate image superposition, based on a grid 
of good reference sources. The positions of the sources selected in 
Section~\ref{data}, together with their uncertainties, were used to compute 
the best transformation needed to superimpose our multi-epoch images. We 
took Observation 3 as a reference. To register the frames, we used a simple 
rotation and translation. We found a strong dependence of the residuals on 
the positions of the reference sources as a function of the distance to the 
aimpoint. Superimposing Observation 1 to Observation 3, these residuals are 
of order 0.2 pixels per coordinate within 4 arcmin from the aimpoint (12 
reference sources); residuals grow to $\sim$0.6 pixels per coordinate between 
4 and 6 arcmin off-axis (8 reference sources); using 10 sources at off-axis 
distances larger than 6 arcmin, the residuals are of order 1.3 pixels per 
coordinate. This is most likely due to the degradation of the PSF with
off-axis angle, which hampers an accurate localization of the sources.

We decided to use only the inner portion of the field (we checked 
{\em a posteriori} that no different astrometric solutions are obtained 
using the entire sample of reference sources). After excluding a source 
deviating at more than $3\sigma$ with respect to the root mean square 
(r.m.s.), we obtained a very good superposition using 11 sources, yielding 
a $1\sigma$ error on the frame registration as small as 50 mas per coordinate. 
The residuals of the reference source positions are $\sim$100 mas per 
coordinate. To be conservative, this value was assumed as the $1\sigma$ 
uncertainty affecting our frame registration. We note that the best fit 
roto-translation implies a frame registration with a similar uncertainty 
(i.e. no significant transformation is required). We repeated the same 
exercise to superimpose Observation 2 to Observation 3, which yielded similar, 
consistent results, although based on a smaller sample of reference sources.
We then computed the target position in the reference frame of Observation 3,
in order to evaluate its possible displacement over the $\sim$5 year interval 
spanned by the observations. 

We found no significant displacement in either coordinate. Accounting for the 
uncertainty in the target position in each image, as well as for the 
uncertainty involved in the frame superposition, a linear fit to the observed 
relative positions sets an upper limit to the proper motion of SGR\,1900+14 of 
17 mas yr$^{-1}$ per coordinate. The $3\sigma$ upper limit to the overall 
proper motion of the source in the plane of the sky is 70 mas yr$^{-1}$.

\section{Absolute astrometry: X-ray versus radio positions.}
A precise localization of SGR\,1900+14 was obtained on 1998 September 10, 
thanks to Very Large Array (VLA) observations of the source after the giant 
flare of August 27 \citep{hurley99}. The accurate radio position was 
$\rm RA = 19^h 07^m 14\fs33$, $\rm Dec.= +09\degr 19\arcmin 20\farcs1$ 
(J2000) with a $1\sigma$ uncertainty of $0\farcs15$ per coordinate
\citep{frail99}. We will take advantage of \cxo\ accurate absolute astrometry
to compare the X-ray position of the target with the 1998 radio position.

\cxo\ absolute localization accuracy for on-axis sources has been carefully 
evaluated by the calibration team. A typical radial uncertainty of 
$\sim$$0\farcs4$ at 68\% confidence level affects ACIS/I 
positions\footnote{See http://cxc.harvard.edu/cal/ASPECT/celmon/.}, 
while for ACIS/S observations this uncertainty is $\sim$$0\farcs2$.
In order to assess (and possibly improve) the absolute astrometry of the 
\cxo\ dataset on SGR\,1900+14, we have cross-correlated the `good' source 
list obtained in Section~\ref{data} with sources in the Two-Micron All-Sky Survey
\citep[2MASS,][]{skrutskie06} catalog, which has an astrometric accuracy of order 
$0\farcs1$.\footnote{See http://spider.ipac.caltech.edu/staff/hlm/2mass/overv/overv.html.}

For Observation~1 and Observation~2, we found 5 coincidences in 2MASS
within $0\farcs8$ of the X-ray position. Increasing the correlation radius 
up to $2\farcs7$ yields no further match. This suggests that the 5 2MASS 
sources are very likely to be the infrared counterparts of the corresponding 
X-ray sources. Infrared colors and X-ray to infrared flux ratio suggest such 
sources to be late stars of K--M spectral class. We used the 5 source
positions to register the \cxo\ images on the accurate 2MASS reference frame 
by fitting a roto-translation, which yielded a r.m.s. of $\sim$200 mas per 
coordinate. 
We note that 
the \cxo-to-2MASS superposition did not require a significant transformation 
(i.e., the corrections are of the same order of the residuals). We repeated 
the same operation using Observation~3, where 2 of the above sources were 
found and were used to adjust the astrometry by fitting a simple translation,
with r.m.s. residual of $\sim$300 mas per coordinate.
We computed the overall uncertainty in the absolute position of SGR\,1900+14
by summing  the target localization accuracy on Chandra images, the
r.m.s. of the Chandra-2MASS frame superposition, and the 2MASS absolute
astrometric accuracy, in quadrature. 

The resulting positions of the target are given in Table \ref{abspos}. These 
positions are (as expected) fully consistent with the accurate radio one 
computed in 1998. With a simple linear fit, we estimate that absolute 
astrometry sets a $3\sigma$ upper limit to the proper motion of SGR\,1900+14 
of 100 mas yr$^{-1}$ per coordinate. Although such a limit is slightly less 
stringent than the one obtained through relative astrometry, this is an 
important consistency check of our result.
\begin{deluxetable}{lllc}
  \tablecolumns{1}
\tablewidth{0pt}
\tablecaption{Multi-epoch absolute positions (J2000) of SGR\,1900+14.\label{abspos}}
\tablehead{
\colhead{RA (error)} & \colhead{Dec. (error)} & \colhead{Date} & \colhead{Instrument}
}
\startdata
$19^{\rm h}07^{\rm m}14\fs33$ ($0\fs15$) & $+09\degr19\arcmin20\farcs1$ ($0\farcs15$) & 1998 Sep 10 & VLA \\
$19^{\rm h}07^{\rm m}14\fs33$ ($0\fs21$) & $+09\degr 19\arcmin 19\farcs6$ ($0\farcs21$) & 2001 Jun 17 & ACIS/I \\
$19^{\rm h}07^{\rm m}14\fs33$ ($0\fs31$) & $+09\degr 19\arcmin 19\farcs8$ ($0\farcs31$) & 2002 Mar 11 & ACIS/S \\
$19^{\rm h}07^{\rm m}14\fs31$ ($0\fs21$) & $+09\degr 19\arcmin 19\farcs8$ ($0\farcs21$) & 2006 Jun 4 & ACIS/I\enddata
\tablecomments{Errors are at 68\% confidence level.}
\end{deluxetable}

\section{Conclusions}
Relative astrometry using a set of three \cxo\ observations spanning 
5 years yields no evidence for any angular displacement of SGR\,1900+14
in the plane of the sky.

Our $3\sigma$ upper limit of 70 mas yr$^{-1}$ 
on the SGR proper motion could still be consistent 
with a physical association of the SGR to G42.8+0.6. Indeed, the $\sim18'$ 
angular separation between the SGR and the SNR center would require, for an
association to hold,
$18' \, / \, (70\, mas \,yr^{-1})\,\sim\,15500$ yr as a $3\sigma$ {\em lower} limit
to the SGR/SNR age. This would point to a system significantly older than usually assumed. 
The age of G42.8+0.6 is rather uncertain, in view of the poorly constrained
distance to the SNR \citep{mlrh01}, and could therefore fit within such a
picture. 
On the other hand, estimating the true age of SGR 1900+14 is very difficult.
The characteristic age $\tau_c \, = \, P/ 2\dot{P}$ of the SGR, derived under standard magneto-dipole 
braking assumptions, is as low as $\sim1300$ yr \citep{woods06}. 
Such a value should be treated with caution, since the spin-down rate of 
SGR 1900+14 has been observed to undergo significant variations \citep[but remaining
very high, in the $6-20\times10^{-11}$ s s$^{-1}$ range, throughout 20 years
of observations,][]{woods06}. The observed $\dot{P}$ changes
did not show any obvious correlation with variability in the SGR persistent 
emission, nor with bursting activity, and ultimately, the physical mechanisms driving the peculiar
SGR spin-down evolution are not understood. 

In any case, it has been argued  \citep{kouveliotou99,thompson00}
that $\tau_c$ could underestimate the true age of the SGR.
For instance, an additional
torque due to a charged particle wind -- leading to a different 
long-term evolution wrt. pure magnetodipole torque (which implies $\dot{P}\propto
P^{-1}$) -- could play an important
role. \citet{thompson00} calculated the braking due to such a wind torque 
(resulting in $\dot{P}\propto P$) and
estimated the true age of the SGR could be of $\sim4000$ years. Even such a 
revised age seems uncomfortably low to be consistent with a $3\sigma$ lower
limit of 15500 yr. An additional hypothesis would be required, namely that we are 
observing SGR 1900+14 in a transient phase of accelerated 
spin-down \citep{kouveliotou99,thompson00}, lasting a fraction $\epsilon$
of the SGR true age $\tau$. Setting $\tau \, = \, \epsilon^{-1} \, P/ \dot{P}$,
our lower limit to the SGR age translates to a $3\sigma$ upper limit to
$\epsilon$ of $\sim0.15$. Unless such a behaviour to be unique
to SGR 1900+14, this would imply that SGRs spend a large majority of
their life in a ``slowly braking'' regime (which ignores the 
braking due to their expected, very high dipole fields). Furthermore, 
slowly braking SGRs should outnumber, by a factor $\epsilon^{-1}$,
SGRs with accelerated braking such as SGR 1900+14 \citep[][proposed to
identify such
slowly braking SGRs with Anomalous X-ray Pulsars; however
AXPs do not display, on average, a significantly slower spin-down than SGRs]{thompson00}.

We believe that it is not worth further speculation to consider these issues. 
In view of the above difficulties in associating SGR 1900+14 and
G42.8+0.6 (taking into account our upper limit to the SGR proper motion), 
and of the 
high chance alignement proability for the two systems, 
Occam's razor argues against any physical link between them.


Such a conclusion adds further support to the association of SGR\,1900+14 with the 
nearby cluster of massive stars discovered by \citet{vrba00}. Morphological 
evidence for a physical interaction of the gas surrounding the SGR and the 
cluster's stars \citep{wachter08} argue against a simple chance alignment 
for the two systems, suggesting that SGR\,1900+14 
originated within the cluster.

Of course, this leaves open the question of the fate of the remnant left by 
the supernova in which the SGR originated. Dense gas and dust clouds
associated with the massive star cluster could hide the SNR emission.
If the birthplace of the SGR lies within the star cluster, a rather 
tight upper limit to its space velocity may be set. The angular separation 
of $\sim$12 arcsec \citep{vrba00} between the position of the SGR and the 
center of the cluster implies a projected velocity of 
$86\,d_{15}\,\tau_{10}^{-1}$ km s$^{-1}$ for the neutron star, where 
$d_{15}$ is the distance in units of 15 kpc and $\tau_{10}$ is the age in 
units of 10 kyr. Unless the SGR velocity is almost aligned along the line
of sight, this estimate is somewhat at odds with one of the basic expectations
of the magnetar model. As discussed by \citet{duncan92}, suppression of 
convection due to the high magnetic field should give rise to anisotropies
in the core-collapse process, resulting in a very high recoil velocity
for the newborn neutron star (\mbox{$\sim$1000 km s$^{-1}$}). We note that to date, no 
evidence for a high space velocity of any magnetar candidate has been found.

In this context, a physical association of SGR\,1806--20 with a nearby cluster 
of massive stars \citep{fuchs99} indirectly becomes more robust. The same is 
true for other magnetar candidates, namely CXO\,J164710.2--455216 
\citep{muno06} and possibly 1E\,1048.1--5937 \citep{gaensler05_1e1048}, 
although a revised distance estimate by \citet{durant06}
argues against the latter association. This mounting 
evidence that a sizeable fraction of the magnetar family had very massive 
progenitors raises several questions concerning magnetar origin and, more 
generally, neutron star formation in core-collapse supernovae. Is there any 
link between progenitor mass and the generation of high magnetic fields in 
their compact remnants? What channels lead to the formation of young neutron
stars as diverse as magnetars 
\citep[B-field$\ \, \sim10^{14}$ G, see e.g. ][]{mereghetti08}, energetic 
`standard' radio pulsars (B-field$\ \, \sim10^{12}$ G) and Central Compact 
Objects \citep[B-field$\ \, \leq10^{11}$ G, see e.g. ][]{deluca08}?
What is the maximum mass for a star to generate a neutron star?
 
Focusing on the case of SGR\,1900+14, more detailed investigations of the 
compact star cluster would be extremely useful, in order to assess the number 
and class of its components and thus to estimate its age. Although this is a 
rather difficult task, in view of the large reddening and the crowded region, 
it would be very rewarding since it would set a lower limit to the mass of 
the progenitor of the soft gamma-ray repeater. 

\acknowledgements
Chandra data analysis is supported by INAF-ASI contract n.I/023/05/0.
KH is grateful for support under the \cxo\ \mbox{AO-2} guest investigator program, 
Smithsonian grant \mbox{GO1-2053X}. This research has made use of data obtained from the Chandra Data Archive 
and software provided by the \cxo\ X-ray Center (CXC) in the application 
packages CIAO, ChIPS, and Sherpa. This publication makes use of data products from the Two-Micron All-Sky 
Survey, which is a joint project of the University of Massachusetts and 
the Infrared Processing and Analysis Center/California Institute of 
Technology, funded by the National Aeronautics and Space Administration 
and the National Science Foundation.

\bibliographystyle{apj}
\bibliography{biblio}

\begin{thebibliography}{}

\bibitem[\protect\citeauthoryear{{De Luca}}{{De Luca}}{2008}]{deluca08}
{De Luca} A.,  2008, in {Bassa} C., {Wang} Z., {Cumming} A., {Kaspi} V.~M.,
  eds., 40 years of pulsars: Millisecond Pulsars, Magnetars and More. Vol.~983
  of AIP Conf. Proc., Melville NY, p.~311

\bibitem[\protect\citeauthoryear{{Duncan} \& {Thompson}}{{Duncan} \&
  {Thompson}}{1992}]{duncan92}
{Duncan} R.~C.,  {Thompson} C.,  1992, \apjl, 392, L9

\bibitem[\protect\citeauthoryear{{Durant} 
\& {van Kerkwijk}}{2006}]{durant06} 
{Durant} M., {van Kerkwijk} M.~H., 2006, \apj, 650, 1070 

\bibitem[\protect\citeauthoryear{{Frail}, {Kulkarni} \& {Bloom}}{{Frail}
  et~al.}{1999}]{frail99}
{Frail} D.~A.,  {Kulkarni} S.~R.,    {Bloom} J.~S.,  1999, \nat, 398, 127

\bibitem[\protect\citeauthoryear{{Fuchs}, {Mirabel}, {Chaty}, {Claret},
  {Cesarsky} \& {Cesarsky}}{{Fuchs} et~al.}{1999}]{fuchs99}
{Fuchs} Y.,  {Mirabel} F.,  {Chaty} S.,  {Claret} A.,  {Cesarsky} C.~J.,
  {Cesarsky} D.~A.,  1999, \aap, 350, 891

\bibitem[\protect\citeauthoryear{{Gaensler}, {Slane}, {Gotthelf} \&
  {Vasisht}}{{Gaensler} et~al.}{2001}]{gaensler01}
{Gaensler} B.~M.,  {Slane} P.~O.,  {Gotthelf} E.~V.,    {Vasisht} G.,  2001,
  \apj, 559, 963

\bibitem[\protect\citeauthoryear{{Gaensler}, {McClure-Griffiths}, {Oey},
  {Haverkorn}, {Dickey} \& {Green}}{{Gaensler}
  et~al.}{2005}]{gaensler05_1e1048}
{Gaensler} B.~M.,  {McClure-Griffiths} N.~M.,  {Oey} M.~S.,  {Haverkorn} M.,
  {Dickey} J.~M.,    {Green} A.~J.,  2005, \apjl, 620, L95

\bibitem[\protect\citeauthoryear{{Hobbs}, {Lorimer}, {Lyne} \&
  {Kramer}}{{Hobbs} et~al.}{2005}]{hobbs05}
{Hobbs} G.,  {Lorimer} D.~R.,  {Lyne} A.~G.,    {Kramer} M.,  2005, \mnras,
  360, 974

\bibitem[\protect\citeauthoryear{{Hurley}, {Li}, {Vrba}, {Luginbuhl},
  {Hartmann}, {Kouveliotou}, {Meegan}, {Fishman}, {Kulkarni}, {Frail}, {Bowyer}
  \& {Lampton}}{{Hurley} et~al.}{1996}]{hurley96}
{Hurley} K. et~al.,  1996, \apjl, 463, L13

\bibitem[\protect\citeauthoryear{{Hurley}, {Cline}, {Mazets}, {Barthelmy},
  {Butterworth}, {Marshall}, {Palmer}, {Aptekar}, {Golenetskii}, {Il'Inskii},
  {Frederiks}, {McTiernan}, {Gold} \& {Trombka}}{{Hurley}
  et~al.}{1999}]{hurley99}
{Hurley} K. et~al.,  1999, \nat, 397, 41

\bibitem[\protect\citeauthoryear{{Hurley}}{{Hurley}}{2000}]{hurley00sgr}
{Hurley} K.,  2000, in {Kippen}, R.~M. and {Mallozzi}, R.~S. and {Fishman},
  G.~J., eds., Gamma-ray Bursts, 5th Huntsville Symposium. Vol.~526 of AIP
  Conf. Proc., Melville NY, p.~763

\bibitem[\protect\citeauthoryear{{Hurley}, {Kouveliotou}, {Duncan}, {Vrba},
  {Garmire}, {Feigelson}, {Woods}, {Gogus} \& {Li}}{{Hurley}
  et~al.}{2002}]{hurley02}
{Hurley} K. et~al.,  2002, MmSAI, 73, 491

\bibitem[\protect\citeauthoryear{{Hurley}, {Boggs}, {Smith}, {Duncan}, {Lin},
  {Zoglauer}, {Krucker}, {Hurford} \& {et~al.}}{{Hurley}
  et~al.}{2005}]{hurley05short}
{Hurley} K. et~al., 2005, \nat, 434,
  1098

\bibitem[\protect\citeauthoryear{{Kaplan}}{{Kaplan}}{2002}]{kaplan02sgr}
{Kaplan} D.~L.,  2002, in {Slane}, P.~O. and {Gaensler}, B.~M. eds., Neutron
  Stars in Supernova Remnants Vol.~271 of ASP Conf. Ser., San Francisco CA,
  p.~266

\bibitem[\protect\citeauthoryear{{Kaplan}, {Kulkarni}, {Frail} \& {van
  Kerkwijk}}{{Kaplan} et~al.}{2002}]{kkf02}
{Kaplan} D.~L.,  {Kulkarni} S.~R.,  {Frail} D.~A.,    {van Kerkwijk} M.~H.,
  2002, \apj, 566, 378

\bibitem[\protect\citeauthoryear{{Kouveliotou} et~al.}{1999}]{kouveliotou99}
{Kouveliotou} C., et al., 1999, \apj, 510, L115 

\bibitem[\protect\citeauthoryear{{Lorimer} \& {Xilouris}}{{Lorimer} \&
  {Xilouris}}{2000}]{lorimer00}
{Lorimer} D.~R.,  {Xilouris} K.~M.,  2000, \apj, 545, 385

\bibitem[\protect\citeauthoryear{{Marsden}, {Lingenfelter}, {Rothschild} \&
  {Higdon}}{{Marsden} et~al.}{2001}]{mlrh01}
{Marsden} D.,  {Lingenfelter} R.~E.,  {Rothschild} R.~E.,    {Higdon} J.~C.,
  2001, \apj, 550, 397

\bibitem[\protect\citeauthoryear{{Mereghetti}}{{Mereghetti}}{2008}]{mereghetti08}
{Mereghetti}, S. 2008, \aapr, 15, 225

\bibitem[\protect\citeauthoryear{{Motch}, {Pires}, {Haberl} \&
  {Schwope}}{{Motch} et~al.}{2007}]{motch07}
{Motch} C.,  {Pires} A.~M.,  {Haberl} F.,    {Schwope} A.,  2007, \apss, 308,
  217

\bibitem[\protect\citeauthoryear{{Motch}, {Pires}, {Haberl}, {Schwope} \&
  {Zavlin}}{{Motch} et~al.}{2008}]{motch08}
{Motch} C.,  {Pires} A.~M.,  {Haberl} F.,  {Schwope} A.,    {Zavlin} V.~E.,
  2008, in {Bassa} C., {Wang} Z., {Cumming} A., {Kaspi} V.~M., eds., 40 years
  of pulsars: Millisecond Pulsars, Magnetars and More. Vol.~983 of AIP Conf.
  Proc., Melville NY, p.~354

\bibitem[\protect\citeauthoryear{{Muno}, {Clark}, {Crowther}, {Dougherty}, {de
  Grijs}, {Law}, {McMillan}, {Morris}, {Negueruela}, {Pooley}, {Portegies
  Zwart} \& {Yusef-Zadeh}}{{Muno} et~al.}{2006}]{muno06}
{Muno} M.~P. et~al.,  2006, \apjl,
  636, L41

\bibitem[\protect\citeauthoryear{Skrutskie et 
al.}{2006}]{skrutskie06} Skrutskie M.~F. et al., 2006, \aj, 131, 
1163 

\bibitem[\protect\citeauthoryear{{Thompson} \& {Duncan}}{{Thompson} \&
  {Duncan}}{1995}]{thompson95}
{Thompson} C.,  {Duncan} R.~C.,  1995, \mnras, 275, 255

\bibitem[{{Thompson} \& {Duncan}(1996)}]{thompson96}
---. 1996, \apj, 473, 322

\bibitem[\protect\citeauthoryear{{Thompson}, {Duncan}, {Woods}, {Kouveliotou},
  {Finger} \& {van Paradijs}}{{Thompson} et~al.}{2000}]{thompson00}
{Thompson} C.,  {Duncan} R.~C.,  {Woods} P.~M.,  {Kouveliotou} C.,  {Finger}
  M.~H.,    {van Paradijs} J.,  2000, \apj, 543, 340

\bibitem[\protect\citeauthoryear{{Vasisht}, {Kulkarni}, {Frail} \&
  {Greiner}}{{Vasisht} et~al.}{1994}]{vasisht94}
{Vasisht} G.,  {Kulkarni} S.~R.,  {Frail} D.~A.,    {Greiner} J.,  1994, \apjl,
  431, L35

\bibitem[\protect\citeauthoryear{{Vrba}, {Henden}, {Luginbuhl}, {Guetter},
  {Hartmann} \& {Klose}}{{Vrba} et~al.}{2000}]{vrba00}
{Vrba} F.~J.,  {Henden} A.~A.,  {Luginbuhl} C.~B.,  {Guetter} H.~H.,
  {Hartmann} D.~H.,    {Klose} S.,  2000, \apjl, 533, L17

\bibitem[\protect\citeauthoryear{{Wachter}, {Ramirez-Ruiz}, {Dwarkadas},
  {Kouveliotou}, {Granot}, {Patel} \& {Figer}}{{Wachter}
  et~al.}{2008}]{wachter08}
{Wachter} S.,  {Ramirez-Ruiz} E.,  {Dwarkadas} V.~V.,  {Kouveliotou} C.,
  {Granot} J.,  {Patel} S.~K.,    {Figer} D.,  2008, \nat, 453, 626

\bibitem[\protect\citeauthoryear{{Woods} \& {Thompson}}{{Woods} \&
  {Thompson}}{2006}]{woods06}
{Woods} P.~M.,  {Thompson} C.,  2006, {in Compact stellar X-ray sources}.
Levin W.~H.~G. and van der Klis M., eds, Cambridge University Press, Cambridge
  UK, p.~547

\end{thebibliography}

\end{document}